# A SPATIAL SCIENTOMETRIC ANALYSIS OF THE PUBLICATION OUTPUT OF CITIES WORLDWIDE

György Csomós
University of Debrecen, Department of Civil Engineering
csomos@eng.unideb.hu

**Abstract**

In tandem with the rapid globalisation of science, spatial scientometrics has become an important research sub-field in scientometric studies. Recently, numerous spatial scientometric contributions have focused on the examination of cities' scientific output by using various scientometric indicators. In this paper, I analyse cities' scientific output worldwide in terms of the number of journal articles indexed by the Scopus database, in the period from 1986 to 2015. Furthermore, I examine which countries are the most important collaborators of cities. Finally, I identify the most productive disciplines in each city. I use GPS Visualizer to illustrate the scientometric data of nearly 2,200 cities on maps. Results show that cities with the highest scientific output are mostly located in developed countries and China. Between 1986 and 2015, the greatest number of scientific articles were created in Beijing. The international hegemony of the United States in science has been described by many studies, and is also reinforced by the fact that the United States is the most important collaborator to more than 75 percent of all cities. Medicine is the most productive discipline in two-thirds of cities. Furthermore, cities having the highest scientific output in specific disciplines show well-defined geographical patterns.

## 1. Introduction

The University of Oxford (England) is one of the most prestigious universities in the world, with a remarkable scientific output[1] (Bonaccorsi, Haddawy, Cicero, & Hassan, 2017; Lin, Huang, & Chen, 2013). It occupies the top position in nearly every university ranking (for example, it is ranked sixth in QS World University Rankings[2] 2016-2017). Tianjin University, the first modern university in both Tianjin and China, occupies a mid-level position in worldwide university rankings (for example, it is only in the 481-490th position in QS World University Rankings 2016-2017), and its scientific output is much smaller than that of the University of Oxford. The University of Oxford and Tianjin University are clearly not on the same tier in terms of scientific output. However, in this spatial scientometric study, I aim to measure and geovisualise scientometric data of cities, not of organisations. The main goal of this paper is to examine cities' scientific output, analyse the characteristics of their international collaboration, and present which scientific disciplines are the most productive ones in each city.

---

[1] Both the definition and the measurement method of scientific output have key importance in this analysis, and will be presented thoroughly in "Data and Methods".
[2] QS World University Rankings 2016-2017: https://www.topuniversities.com/university-rankings/world-university-rankings/2016

Using the aforementioned example as an illustration, I aim to compare the scientific output of the cities of Oxford and Tianjin; i.e., the scientific output of a mid-sized city with fewer than 170,000 inhabitants against the scientific output of a megacity with more than 15 million people.

According to Frenken, Hardeman, & Hoekman (2009), the first studies discussing the spatial distribution of science were published in the 1970s; however, spatial scientometrics has only recently begun to attract more attention (Bornmann & Waltman, 2011). The geospatial measurement of cities' (entire urban regions') scientific output using various scientific indicators appears in many studies. Matthiessen & Schwarz (1999) examine the scientific output of European urban regions in terms of the number of papers in the Science Citation Index (SCI). Zhou, Thijs & Glänzel (2009a) analyse the scientific output of Chinese (provincial-level) administrative regions in terms of the number of scientific publications collected in the SCI Expanded database. In an article published in *Nature News*, van Noorden (2010) discusses which urban regions produce the best research in the world, and whether their success can be replicated. Bornmann, Leydesdorff, Walch-Solimena, & Ettl (2011) and Bornmann & Waltman (2011) classify cities and urban regions as "centers of excellence in scientific research", based on the total number of excellent papers (top 1% most highly cited papers). Bornmann & Leydesdorff (2011), and Bornmann & Leydesdorff (2012) examine whether there is a relationship between the total scientific output of cities and the number of highly cited papers published in those cities. Csomós & Tóth (2016) and Csomós (2017) explore the global position of cities in terms of corporate research and development, based on the number of scientific publications created by corporate researchers and engineers.

These studies are limited by their focus on specific geographic regions (e.g., Europe, China, etc.) or specific research areas (e.g., neuroscience, physics and astronomy, etc.). In this paper, I examine cities' scientific output based on Scopus data, in a search for answers to three research questions:
- Which cities in the world have the highest scientific output, and how has this output changed over time?
- For a given city, which countries are its most important collaborators?
- Which disciplines are the most productive in each city?

The structure of the paper is as follows: In Section 2, I describe the data collection process and methodology. In Section 3, a parallel is made between the results of the study and the answers to the research questions. In Section 4, I present the limitations of the analysis; and finally, in Section 5, I present the conclusions.

## 2. Data and methodology

### 2.1. Data collection

Before presenting the results, I will address two important issues that need to be clarified: First, how can scientific output of cities be measured, and second, which cities should be examined in terms of scientific output?

#### 2.1.1. Determination of the source and type of scientific publications

In this analysis, cities' scientific output is obtained by measuring the total number of publications written by authors who are affiliated with a professional organization (e.g., universities, firms, hospitals, governmental and non-governmental institutions, etc.) in that city. Scientific publication data is indexed by several databases. To achieve the main goals of this spatial scientometric analysis, Web of Science and Scopus are chosen as the most relevant databases containing scientific publication data. Both databases include the authors' affiliations (i.e., names, addresses) and sum up the total number of articles created by authors affiliated with specific organisations in the same city. Furthermore, both databases assign a publication to the cities in which they were produced, rather than the cities in which the affiliated organisations' headquarters are located[3]. This is an important distinction, because some large organisations (universities, multinational corporations, national or international research institutions) house their semi-autonomous but scientifically productive units (university hospitals, corporate research centres) in one city and their command and control centres in another city (corporate headquarters, universities' main campuses, etc.). For example, IBM, a corporation that is among the world's leaders in scientific output, is headquartered in Armonk, New York, but only 17 percent of the corporation's total number of publications were created here (i.e., the authors have indicated this location as the institutional address on their publication). Scopus assigns 18 additional affiliations to IBM, including the Thomas J. Watson Research Center in Yorktown Heights, New York, where 47 percent of IBM's publications were created. As a matter of fact, both Armonk and Yorktown Heights are located within the New York-Newark metropolitan area, whereas the IBM Almaden Research Center, which has roughly the same number of publications as the Armonk headquarters, is located in San Jose, California. Other scientifically productive IBM subsidiaries are in Ruschlikon, (Switzerland), Bangalore (India), Haifa (Israel), Tokyo, and Beijing. Obviously, the number of scientific publications created in subsidiary cities does not increase the scientific output of the city which is home to the corporation's headquarters, not even if research activities (which generally come before the process of publishing) are controlled from the headquarters.

In this work, the Scopus database was chosen as the main source of scientific publication data[4]. Although there is a significant overlap between the contents of Scopus and the Web of Science (Norris & Oppenheim, 2007), two factors made Scopus stand out: 1) a greater number of journals, and a much greater number of non-English-language journals are indexed by Scopus, than by the Web of Science (Li, Qiao, Li, & Jin, 2014; Mongeon & Paul-Hus, 2016; Vieira & Gomes, 2009); 2) Scopus was technically easier to manage and data mine. The main advantage of Scopus is that when typing the name of a country (e.g., Sweden) into the "Affiliation search" field, the system lists all affiliations in that country (e.g., 395 affiliation results in Sweden) and assigns them to cities (e.g., 75 cities in Sweden). In contrast, this function is only partially supported in the Web of Science, which means that after typing the name of a country into the "Address search" field, a list of organisations will appear, but a list of cities will not be available. In order to create a list of cities, affiliation addresses (Organizations-Enhanced) must be checked one by one. In the case of Sweden, for example, the number of affiliations in the Web of Science exceeds 500.

---

[3] It should be noted that both Web of Science and Scopus present the addresses (i.e., the name of cities) reported by the authors of a publication. In most cases, these are the addresses at which a publication was produced. However, it is generally at the authors' discretion to choose what address they wish to indicate on a publication, and if they choose to report the address of their organization's headquarters, then this will be the information that is presented in Web of Science and Scopus.

[4] It should be noted that Scopus coverage before 1996 is traditionally less extensive than its coverage after 1996; however, Elsevier has made a commitment to further expand its coverage of pre-1996 publications and citations (Elsevier, 2014; Harzing & Alakangas, 2016). Naturally, this fact may affect the findings of this analysis, leastwise that of the first decade (1986-1995).

The following constraints were implemented in order to improve the objectivity of the study: 1) only journal articles will be selected for the analysis; and 2) the data will only include papers produced between 1986 and 2015.

The reason for constraint (1) is that journal articles are generally considered as the most prestigious of scientific publications, since they are "the basic means of communicating new scientific knowledge" (Braun, Glänzel, & Schubert, 1989: 325). In fact, approximately 62 percent of the total number of publications indexed by Scopus are journal articles (Elsevier, 2016). Therefore, I excluded all other types of publications (e.g., conference papers, reviews, letters, book chapters, editorials, notes, reports, etc.).

The reason for constraint (2) is that, by collecting data from a broad span of 30 years, the results will be more objective and balanced. Scopus includes articles published in developed countries (primarily in the United States, the United Kingdom, France, Germany, Japan) dating back to 1823; however, the research activities of developing countries[5] did not begin until the second half of the 20$^{th}$ century (see, for example, Gao, Guo, Sylvan, & Guan, 2010; Leta & De Meis, 1996; Moin, Mahmoudi, & Rezaei, 2005; Panat, 2014; Zhou, Thijs, & Glänzel, 2009a). Consequently, Scopus includes only a small amount of scientific publications created before the 1970s and 1980s. To illustrate, authors affiliated with the United Kingdom generated 78.8 percent of its articles in the period between 1986 and 2015, and Scopus includes articles that were published as far back as 1858. However, authors with Chinese affiliation produced only 0.28 percent of its articles before 1986, with the oldest published in 1909.

In addition, by using a 30-year period, total scientific output can be examined in decades (1986-1995, 1996-2005, and 2006-2015), revealing global tendencies and trends.

**2.1.2. Selection methodology of cities**

Having determined the database, the type of publication, and the publishing period of the scientific contributions in this analysis, I will now introduce the method for measuring cities' scientific output. First, however, I would like to discuss which cities should be examined in terms of scientific output.

Cities that house prestigious universities (e.g., Boston, Oxford, Cambridge, England, and Cambridge, Massachusetts) have high scientific output; this is well-known. In fact, the same can be said for any major world city (e.g., New York, Los Angeles, London, Paris, Tokyo), because, they are home to leading universities, national and international research institutions, and enjoy favourable benefits due to their global economic status. Studies by Matthiessen & Schwarz (1999), van Noorden (2010), Bornmann & Leydesdorff (2011), Bornmann & Waltman (2011), and Bornmann, Leydesdorff, Walch-Solimena, & Ettl (2011) also confirm the unique position held by large cities in international science. According to other scholars, leading multinational companies (MNCs), especially those which operate in research-intensive industries, also publish a great number of scientific articles (Chang, 2014; Csomós & Tóth, 2016; Csomós, 2017; Godin, 1996; Halperin & Chakrabarti, 1987; Hicks, Ishizuka, Keen, & Sweet, 1994; Hicks, 1995). MNCs are generally far more diversified than universities; this is likely due to MNCs' trend of housing their research facilities in the most innovative environments possible, in order to improve their organisation's ability to leverage knowledge (D'Agostino & Santangelo, 2012; Gerybadze & Reger, 1999; Pearce, 1999; Zander, 1999). Naturally, the "most innovative environment" for MNCs are university cities[6] and world cities, but the basis for the localisation of corporate research centres, and their research-oriented subsidiaries, includes several business factors that often overshadow the needs of scientific staff.

---

[5] There are debates surrounding whether China is still a developing country. According to the World Bank, China's per capita nominal GDP is relatively low, and because of this fact, the country remains a developing country (see, The World Bank, Overview: http://www.worldbank.org/en/country/china/overview). However, others say that as a result of the rapid development the country has undergone in the last decade, China can also be considered a developed country (see, for example, ChinaPower: http://chinapower.csis.org/is-china-a-developed-country, and Foreign Policy: http://foreignpolicy.com/2014/09/25/is-china-still-a-developing-country).

There are also some megacities that produce a large number of scientific articles (e.g., Cairo, São Paulo, Tehran, and New Delhi). These cities have an extremely large population and area, and consequently, they host many organisations that have scientific output in both small and large measure. For example, Scopus includes 139 organisations affiliated with Boston, one of the top scientific centres in the world, but 202 organisations are affiliated with New Delhi, a megacity.

Consequently, instead of choosing specific cities or a group of cities to be examined, the opposite had to be done. Cities having the highest scientific output (definition to come) were selected and examined. This procedure made it necessary to scrutinise the entire Scopus database, assigning journal articles to the cities in which the corresponding articles were created. These cities are located in 232 countries and territories (as defined by the United Nations), and, in total, there are 52,577 organisations affiliated with them. In the analysis, only the cities that had at least 1,000 journal articles indexed by Scopus in the period from 1986 to 2015 were included. These criteria were met by 2,194 cities[7].

## 2.2. Methods

In this paper, cities' scientific output is determined from the total number of articles indexed by Scopus in the period from 1986 to 2015. No other selection criterion is used. In the Scopus database, authors' affiliations, and thus the articles they published, are assigned to specific cities; however, in some cases, results are quite confusing or misleading. Scopus sometimes associates more than one city to a single name, even if those cities that share a name are located in different countries. For example, there are 176 organizations affiliated with Melbourne. The affiliation does not discern that there are three distinct cities (located in Australia, Canada, and the United States) that share the name Melbourne. The separation of these cities is very easy, because Scopus allows users the ability to refine results. It is far more problematic if there are cities who share the same name within a single country. In the United States, some cities share the same name but are in a different state. The problem that arises is based on the query method. Scopus can refine a search by cities and countries, but not by states. For example, Scopus assigns 70 organisations to Columbia (United States); however, after more careful review, the results show that these organisations are affiliated with three different cities in three different states: Columbia, Maryland; Columbia, Missouri; and Columbia, South Carolina. When this problem is identified, the address of each affiliation must be investigated one by one (see, for example, Arlington, Virginia/Texas, or Springfield, Illinois/Missouri/Massachusetts).

---

[6] Tödtling (1994) and Owen-Smith & Powell (2004) present this phenomenon through the example of Boston.
[7] Some affiliations are not located in cities. For example, the Moffett Federal Airfield is an airport that is in an unincorporated part of Santa Clara County between Mountain View and Sunnyvale, California. Another example is the Research Triangle Park, one of the world's largest research parks, which lies between Durham, Raleigh and Chapel Hill, North Carolina. These type of locations are treated as "cities" in the analysis, as if they were autonomous administrative units. Some affiliations encompass more than one settlements (cities, towns, villages, etc.), and technically it is not possible to separate them. For example, Japanese towns and villages belong to districts which are used primarily in the Japanese addressing system and to identify the relevant geographical areas. In this case, instead of towns and villages, Scopus indicates the name of the districts.

I use GPS Visualizer's (www.gpsvisualizer.com) "plot data points", a free online application, to illustrate cities' scientific output on maps (for more information on using the software for spatial scientometric studies, see Bornmann, Leydesdorff, Walch-Solimena, & Ettl, 2011; Bornmann & Waltman, 2011; Leydesdorff & Persson, 2010; Waltman, Tijssen, & Eck, 2011). I identified the cities' coordinates using LatLong.net (www.latlong.net).

In my analysis, the cities' total scientific output in terms of the number of published scientific articles is much higher than the total number of articles indexed by Scopus. The reason for this is as follows: Scopus indexes article titles, and in this analysis, an article can belong to multiple cities depending on the affiliations of the authors. This especially characterises articles focused on natural sciences (e.g., physics, chemistry, and mathematics) and life sciences, as pointed out by Adams (2012); Castelvecchi (2015); Glänzel (2001); Hsu & Huang (2009); and Huang (2015).

## 3. Results

### 3.1. Cities' scientific output

This analysis contains information on the scientific output of 2,194 cities with at least 1,000 journal articles indexed by Scopus between 1986 and 2015. The greatest number of cities that fulfilled the criteria above are in the United States (i.e., 21.8 percent of all cities), and the greatest number of articles were also created in cities in the United States (as shown in Table 1). Japan is ranked second for the number of cities that fulfilled the criteria; however, the second largest number of articles was published in China. This means that Chinese cities generally produce more articles than do Japanese cities, which is also reflected by the fact that China is ranked second in terms of the output per city (Singapore, a sovereign city-state, is first in this ranking). Chinese cities generally have huge size and population, but they are not surrounded by suburbs, mostly because suburbs are annexed by adjacent cities. Conversely, in developed countries (e.g., Australia, Canada, Japan, and the United States), major cities often have large suburban areas around them, and in some cases, the suburbs produce more scientific articles than does the central city. For example, Tokyo is the central city of the Greater Tokyo Area. It is surrounded by more than 30 suburbs that have at least 1,000 articles indexed by Scopus, while Beijing (a city similar in size to the Greater Tokyo Area) does not have any suburbs. Furthermore, in China, only a small, albeit growing, number of cities have been involved in national research activities that may lead to scientific publications (Andersson, Gunessee, Matthiessen, & Find, 2014; Grossetti, Eckert, Gingras, Jégou, Larivière, & Milard, 2014; He, Zhang, & Teng, 2005; Zhou, Thijs, & Glänzel, 2009a).

Most cities with the highest scientific output are located in two types of countries: 1) countries with a very large area and lots of cities (e.g., the Russian Federation, the United States, China, Brazil, and India); and 2) mid-sized developed countries with high population density (e.g., Japan, France, Germany, the United Kingdom, Italy, Spain, South Korea, the Netherlands, and Sweden). Therefore, measuring the total number of articles published in all cities of a country is not an advantageous measurement for every country. For example, India is ranked second in terms of the number of cities fulfilling the criteria, but is ranked 10[th] in terms of the total number of articles. India's average of 7,677 articles per city results in its 59[th] position ranking. China and the United Kingdom, on the other hand, show an opposite pattern. The United Kingdom occupies the seventh position in the ranking of cities that fulfil the criteria, but the fourth position both in terms of the total number of articles and the number of articles per city.

**Table 1. Top 20 countries that are home to cities having high scientific output**

| Rank | Country | Total number of cities | Rank | Country | Total number of articles in 1986-2015 | Rank | Country | Number of articles per city |
|---|---|---|---|---|---|---|---|---|
| 1 | United States | 478 | 1 | United States | 9,984,536 | 1 | Singapore | 140,915 |
| 2 | Japan | 151 | 2 | China | 3,757,538 | 2 | China | 34,159 |
| 3 | India | 122 | 3 | Japan | 2,703,448 | 3 | Austria | 26,847 |
| 4 | France | 113 | 4 | United Kingdom | 2,386,783 | 4 | United Kingdom | 25,943 |
| 5 | China | 110 | 5 | Germany | 2,159,851 | 5 | Finland | 24,017 |
| 6 | Germany | 97 | 6 | France | 1,632,996 | 6 | Canada | 23,286 |
| 7 | United Kingdom | 92 | 7 | Italy | 1,351,816 | 7 | Switzerland | 23,049 |
| 8 | Italy | 69 | 8 | Canada | 1,280,713 | 8 | Netherlands | 22,446 |
| 9 | Brazil | 60 | 9 | Spain | 975,930 | 9 | Germany | 22,267 |
| 10 | Spain | 57 | 10 | India | 936,550 | 10 | Israel | 21,897 |
| 11 | Canada | 55 | 11 | Australia | 923,258 | 11 | Australia | 21,471 |
| 12 | Turkey | 45 | 12 | South Korea | 813,683 | 12 | United States | 20,888 |
| 13 | Australia | 43 | 13 | Netherlands | 740,704 | 13 | South Korea | 20,864 |
| 14 | South Korea | 39 | 14 | Russian Federation | 703,048 | 14 | Belgium | 20,680 |
| 15 | Russian Federation | 39 | 15 | Brazil | 675,918 | 15 | Italy | 19,592 |
| 16 | Iran | 36 | 16 | Sweden | 504,717 | 16 | Russian Federation | 18,027 |
| 17 | Netherlands | 33 | 17 | Switzerland | 437,937 | 17 | Japan | 17,904 |
| 18 | Sweden | 29 | 18 | Turkey | 411,007 | 18 | New Zealand | 17,740 |
| 19 | Mexico | 29 | 19 | Poland | 398,995 | 19 | Denmark | 17,644 |
| 20 | Poland | 23 | 20 | Belgium | 351,558 | 20 | Ireland | 17,446 |
|  | Other countries | 474 |  | Other countries | 4,935,306 |  |  |  |

The Scopus data allows us to determine which countries are home to the cities with the highest scientific output in the world. Fig. 1 shows that a vast majority of cities with high scientific output are located in three major geographic regions: 9.7 percent of the total number of cities can be found in North America, especially on the East Coast of the United States; 13.7 percent in East Asia, and 20.8 percent in Western Europe. Furthermore, an increasing number of cities are located on the West Coast of the United States (primarily in California), in the southeast region of Brazil, on the East Coast of Australia, in the western region of Africa, and in India.

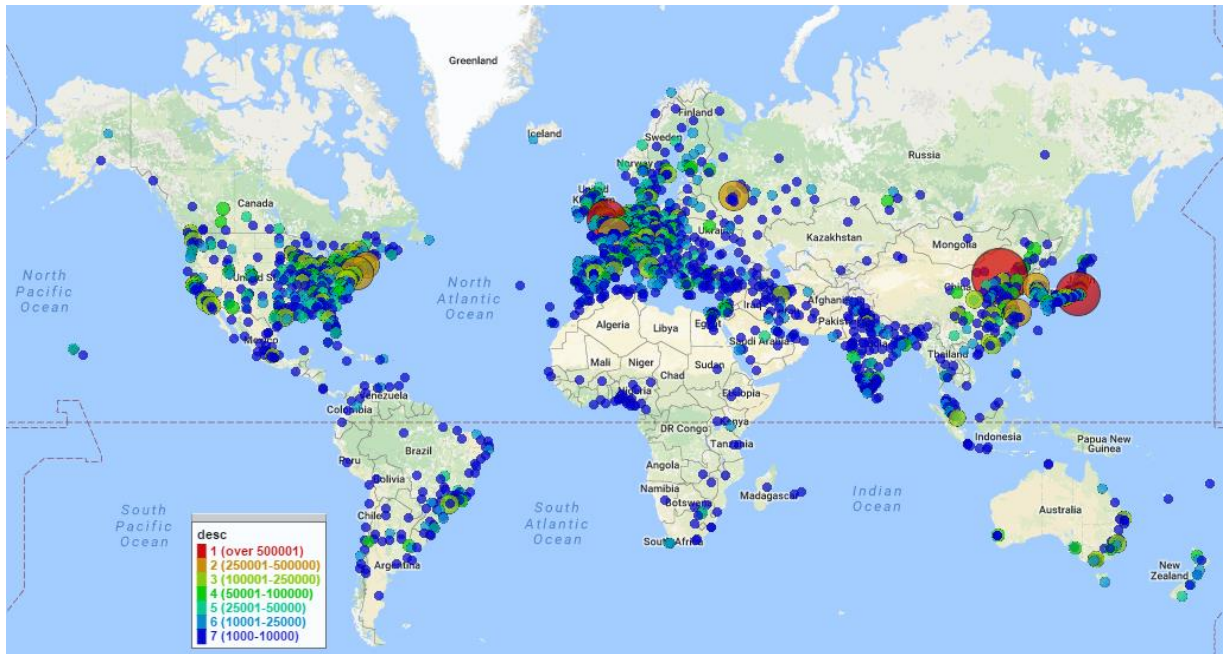

**Fig. 1. Location of cities having the highest scientific output, 1986-2015**
**(A zoomable version of this map is available at https://dea.lib.unideb.hu/dea/handle/2437/241236)**

In the period from 1986 to 2015, authors affiliated with Beijing produced the highest number of articles. The scientific output of Beijing exceeded that of Tokyo, which was second in the rankings, by 34 percent. The Chinese capital had 91 percent more articles than did New York City (Table 2). The massive growth of Beijing's scientific output has been observed and described in past studies (Andersson, Gunessee, Matthiessen, & Find, 2014; Hu, Guo, & Hou, 2017; Maisonobe, Eckert, Grossetti, Jégou, & Milard, 2016; van Noorden, 2010; Zhou, Thijs, & Glänzel, 2009a); however, the growth rate of the output is quite astonishing. In the period from 1986 to 2015, a greater number of articles were created in Beijing than in Spain. This is almost the same amount produced by the entire Eastern European region (not including the Russian Federation). The amount of Beijing's scientific output is graphically illustrated by the fact that a greater number of articles were created in the Chinese capital in the 10-year period from 2006 to 2015 than in Tokyo in a 30-year period. Zhou, Thijs, & Glänzel (2009a) state, however, that Beijing's total scientific output in proportion to China has been gradually decreasing since the early 2000s, primarily because of the increasing scientific output of some emerging mainland Chinese cities; including Shanghai, Nanjing, Wuhan, Xi'an, Hong Kong, and Guangzhou; all of which have been growing at a swift rate[8]. Nonetheless, this progress does not influence the global position of Beijing, since the productivity of other major cities (e.g., Tokyo, London, Seoul, Paris, and New York City) in terms of the total scientific output has dropped far behind. The explosive strengthening of Beijing's power in the global economy and finance has been known for a while (see, for example, Chen & Chen, 2015; Csomós, 2013; Lai, 2012; Pan, Guo, Zhang, & Liang, 2015; Timberlake, Wei, Ma, & Hao, 2014), but it has recently been observed that the city now occupies a leading position in international science[9].

---

[8] Naturally, one might suggest that this rapid growth in terms of total scientific output of Chinese cities reflects the fact that Scopus has been indexing an increasing number of Chinese-language journals (see, for example, Li, Qiao, Li, & Jin, 2014; Lin & Zhang, 2007). Scopus includes more than 34,000 journal titles, from which about 19,100 journals publish articles in English, but fewer than 400 journals publish articles in Chinese (30 of them are bilingual). The number of Chinese-language journals in Scopus is insignificant as compared to that of English-language journals. It is also important to note that only one-third of the Chinese-language journals have been included in Scopus since 2006 (i.e., since the beginning of the third decade of the current study), but two-thirds of them have been included since 1978 (see data in Scopus Source List, https://www.elsevier.com/solutions/scopus/content). The increase in the total scientific output of Chinese cities between 2006 and 2015 is due to the fact that Chinese scholars are publishing more and more articles in English-language journals; not that Scopus is indexing a growing number of Chinese-language journals.

**Table 2. The world's top 100 cities having the highest scientific output, 1986-2015**

| Rank | City | Country | Number of articles, 1986-2015 | Rank | City | Country | Number of articles, 1986-2015 |
|---|---|---|---|---|---|---|---|
| 1 | Beijing | China | 859,616 | 51 | Ottawa | Canada | 121,043 |
| 2 | Tokyo | Japan | 642,608 | 52 | Madison | United States | 119,388 |
| 3 | London | United Kingdom | 561,242 | 53 | San Francisco | United States | 116,834 |
| 4 | New York City | United States | 450,863 | 54 | Suita | Japan | 116,733 |
| 5 | Paris | France | 447,577 | 55 | Minneapolis | United States | 115,694 |
| 6 | Moscow | Russian Federation | 382,627 | 56 | Tsukuba | Japan | 115,254 |
| 7 | Shanghai | China | 340,746 | 57 | Copenhagen | Denmark | 114,019 |
| 8 | Seoul | South Korea | 326,055 | 58 | Brisbane | Australia | 112,970 |
| 9 | Boston | United States | 320,946 | 59 | Warsaw | Poland | 110,447 |
| 10 | Washington D.C. | United States | 240,822 | 60 | Manchester | United Kingdom | 108,270 |
| 11 | Rome | Italy | 239,861 | 61 | Vancouver | Canada | 106,990 |
| 12 | Los Angeles | United States | 224,614 | 62 | Mexico City | Mexico | 105,839 |
| 13 | Nanjing | China | 215,829 | 63 | Nagoya | Japan | 105,498 |
| 14 | Madrid | Spain | 213,385 | 64 | Daejeon | South Korea | 104,781 |
| 15 | Philadelphia | United States | 206,482 | 65 | Ankara | Turkey | 103,406 |
| 16 | Chicago | United States | 200,583 | 66 | Sendai | Japan | 103,146 |
| 17 | Toronto | Canada | 199,941 | 67 | St. Louis | United States | 102,778 |
| 18 | Cambridge | United States | 191,446 | 68 | New Delhi | India | 102,676 |
| 19 | São Paulo | Brazil | 190,171 | 69 | Changsha | China | 101,948 |
| 20 | Baltimore | United States | 187,939 | 70 | Prague | Czech Republic | 101,884 |
| 21 | Bethesda | United States | 180,897 | 71 | Gainesville | United States | 101,205 |
| 22 | Berlin | Germany | 180,103 | 72 | Columbus | United States | 100,791 |
| 23 | San Diego | United States | 176,475 | 73 | Tianjin | China | 100,713 |
| 24 | Montreal | Canada | 171,847 | 74 | New Haven | United States | 98,175 |
| 25 | Wuhan | China | 170,159 | 75 | Heidelberg | Germany | 97,000 |
| 26 | Sydney | Australia | 169,334 | 76 | Palo Alto | United States | 96,196 |
| 27 | Munich | Germany | 168,183 | 77 | Harbin | China | 95,769 |
| 28 | Houston | United States | 167,057 | 78 | Durham | United States | 95,561 |
| 29 | Xi'an | China | 163,594 | 79 | Helsinki | Finland | 95,341 |
| 30 | Barcelona | Spain | 163,297 | 80 | Cleveland | United States | 94,230 |
| 31 | Kyoto | Japan | 162,520 | 81 | Uppsala | Sweden | 92,684 |
| 32 | Hong Kong | China | 160,863 | 82 | Utrecht | Netherlands | 92,181 |
| 33 | Seattle | United States | 159,926 | 83 | Edinburgh | United Kingdom | 89,279 |
| 34 | Tehran | Iran | 156,096 | 84 | Davis | United States | 86,827 |
| 35 | Stockholm | Sweden | 155,417 | 85 | Fukuoka | Japan | 86,399 |
| 36 | Cambridge | United Kingdom | 153,285 | 86 | Austin | United States | 86,119 |
| 37 | Amsterdam | Netherlands | 148,474 | 87 | Chapel Hill | United States | 85,587 |
| 38 | Atlanta | United States | 148,040 | 88 | Athens | Greece | 85,122 |
| 39 | Ann Arbour | United States | 144,558 | 89 | Hefei | China | 84,066 |
| 40 | Milan | Italy | 143,404 | 90 | Leuven | Belgium | 83,874 |
| 41 | Singapore | Singapore | 140,915 | 91 | Glasgow | United Kingdom | 83,315 |
| 42 | Guangzhou | China | 139,952 | 92 | Brussels | Belgium | 81,373 |
| 43 | Melbourne | Australia | 137,475 | 93 | Shenyang | China | 80,870 |
| 44 | Oxford | United Kingdom | 135,777 | 94 | Edmonton | Canada | 80,261 |
| 45 | Berkeley | United States | 135,466 | 95 | Sapporo | Japan | 80,191 |
| 46 | Pittsburgh | United States | 135,307 | 96 | State College | United States | 79,919 |
| 47 | Zürich | Switzerland | 130,800 | 97 | Tucson | United States | 79,893 |
| 48 | Hangzhou | China | 130,555 | 98 | Ithaca | United States | 79,317 |
| 49 | Chengdu | China | 128,612 | 99 | Budapest | Hungary | 78,366 |
| 50 | Vienna | Austria | 124,099 | 100 | Tel Aviv-Yafo | Israel | 77,245 |

In the period between 1986 and 1995, cities that produced the highest scientific output were located in developed countries (Table 3). In this period, cities from developing countries had quite low scientific output. For example, the combined output of all Chinese cities did not reach the value of London. In the first decade of this period, only Beijing produced considerable scientific output. New Delhi (with 13,332 articles) was the next city on the list from a developing country, but only occupied the 92$^{nd}$ position, behind Pasadena, California. In tandem with the Chinese economic reforms initiated in 1978, the nation's innovation system has also been developed (Liu & White, 2001; Liu, Simon, Sun, & Cao, 2011). In that, the scientific output of universities and the Chinese Academy of Sciences have grown rapidly; moreover, companies have gradually been getting involved in the production of science (Csomós & Tóth, 2016). In the period from 2006 to 2015, eight Chinese cities produced more than 100,000 articles of the top 25 cities worldwide, while in the United States, there were only four such cities. In terms of volume, similar economic reforms occurred in South Korea, but they did not result in the geographical expansion of science, except for in Seoul and Daejeon. In fact, other cities in South Korea are only marginal in the national innovation system (Shapiro & Park, 2012). Notably, however, in the period from 1986 to 1995, Seoul occupied the 138$^{th}$ position in the ranking of cities with the highest scientific output, but between 2006 and 2015, it climbed to the 5$^{th}$ position.

---

[9] Based on total scientific output, Beijing competes with European and Northern American cities; however, according to Andersson, Gunessee, Matthiessen, & Find (2014: 2969), Beijing "occupies a peripheral position in the production of new scientific breakthroughs."

According to Leta, Glänzel, & Thijs (2006: 87), "the development of scientific and technological infrastructure and the formation and expansion of Brazilian academic community" started in the 1960s. These actions have primarily targeted that country's most industrialised city, São Paulo, which is also home to the leading Brazilian universities (de Almeida & Guimarães, 2013). In the first decade of the study period, 28.1 percent of Brazil's scientific output came from São Paulo, which increased to 35.6 percent by 2006-2015; i.e., the position of São Paulo in Brazilian science has significantly strengthened.

**Table 3. Cities having the highest scientific output by decades**

| Rank | City | Country | Number of articles, 1986-1995 | City | Country | Number of articles, 1996-2005 | City | Country | Number of articles, 2006-2015 |
|---|---|---|---|---|---|---|---|---|---|
| 1 | Tokyo | Japan | 139,268 | Tokyo | Japan | 227,137 | Beijing | China | 664,414 |
| 2 | London | United Kingdom | 124,099 | London | United Kingdom | 169,748 | Tokyo | Japan | 276,203 |
| 3 | New York City | United States | 95,011 | Beijing | China | 164,681 | London | United Kingdom | 267,395 |
| 4 | Paris | France | 90,075 | Moscow | Russian Federation | 145,272 | Shanghai | China | 262,635 |
| 5 | Moscow | Russian Federation | 64,584 | New York City | United States | 139,917 | Seoul | South Korea | 239,438 |
| 6 | Boston | United States | 60,513 | Paris | France | 137,513 | Paris | France | 219,989 |
| 7 | Los Angeles | United States | 51,373 | Boston | United States | 92,433 | New York City | United States | 215,935 |
| 8 | Washington D.C. | United States | 49,433 | Seoul | South Korea | 76,345 | Nanjing | China | 176,284 |
| 9 | Bethesda | United States | 47,155 | Washington D.C. | United States | 75,493 | Moscow | Russian Federation | 172,771 |
| 10 | Philadelphia | United States | 46,167 | Rome | Italy | 72,439 | Boston | United States | 168,000 |
| 11 | Chicago | United States | 45,028 | Los Angeles | United States | 66,309 | Tehran | Iran | 142,180 |
| 12 | Rome | Italy | 40,968 | Shanghai | China | 65,065 | São Paulo | Brazil | 135,257 |
| 13 | Cambridge | United States | 37,068 | Philadelphia | United States | 62,153 | Wuhan | China | 134,840 |
| 14 | Toronto | Canada | 36,730 | Madrid | Spain | 61,977 | Xi'an | China | 131,360 |
| 15 | Baltimore | United States | 35,708 | Berlin | Germany | 58,309 | Rome | Italy | 126,454 |
| 16 | San Diego | United States | 35,631 | Chicago | United States | 58,189 | Madrid | Spain | 123,605 |

| 17 | Houston | United States | 34,680 | San Diego | United States | 57,871 | Guangzhou | China | 116,952 |
| --- | --- | --- | --- | --- | --- | --- | --- | --- | --- |
| 18 | Kyoto | Japan | 33,458 | Bethesda | United States | 57,445 | Washington D.C. | United States | 115,896 |
| 19 | Munich | Germany | 33,152 | Cambridge | United States | 56,905 | Toronto | Canada | 107,079 |
| 20 | Montreal | Canada | 32,699 | Kyoto | Japan | 56,837 | Los Angeles | United States | 106,932 |
| 21 | Berlin | Germany | 31,580 | Toronto | Canada | 56,132 | Hangzhou | China | 105,103 |
| 22 | Stockholm | Sweden | 30,791 | Baltimore | United States | 55,986 | Chengdu | China | 105,099 |
| 23 | Seattle | United States | 30,603 | Hong Kong | China | 53,595 | Barcelona | Spain | 102,508 |
| 24 | Beijing | China | 30,521 | Munich | Germany | 50,603 | Sydney | Australia | 98,556 |
| 25 | Ann Arbour | United States | 29,694 | Montreal | Canada | 50,225 | Philadelphia | United States | 98,162 |

Tehran is unique in terms of the growth rate of its scientific output. It occupies the 34th position (see Table 2), but 91 percent of its total number of articles were created between 2006 and 2015 (see Table 3). If we compare this result with the previous decade, Tehran's scientific output grew by almost 1,100 percent, which was the highest growth rate among leading cities (the growth rate of Beijing was only 403 percent). The reason for Tehran's (and Iran's, respectively) very low scientific output from the 1980s can largely be attributed to the negative effects of the Iran-Iraq War of 1980-1988. Moin, Mahmoudi, & Rezaei (2005) claim that during the war, Iran's scientific output dramatically decreased. The amount of the scientific output it achieved in the 1970s could only be reached by the end of the 1990s. Rapid growth began in the mid-2000s, due to increasing scientific productivity in the field of medicine (Abolghassemi Fakhree & Jouyban, 2011).

**3.2. The most important international collaborators**

The Scopus "search results analyser" makes it possible to examine the characteristics of cities' international collaboration; i.e., for a given city (more precisely, authors affiliated with that city), which country (more precisely, authors affiliated with that country) is the most important collaborator. For example, in the period between 1986 and 2015, 320,946 articles that were created in Boston were co-written with authors from more than 160 countries. Of them, 16,638 co-authors were from the United Kingdom, 16,562 co-authors from Germany, 15,136 co-authors from Canada, 10,052 co-authors from France, etc. Considering these facts, it can be determined that the most important collaborator of Boston is the United Kingdom.

     The hegemony of the United States in international science is a well-known phenomenon[10] (Paasi, 2005). This is also underpinned by the fact that the United States is the most important collaborator for almost every country in the world, and in almost every discipline (see, for example, He & Guan, 2008; Liu, Hu, Tang, & Wang, 2015; Lu & Wolfram, 2010; Maisonobe, Eckert, Grossetti, Jégou, & Milard, 2016; Zitt, Bassecoulard, & Okubo, 2000). Table 4 shows that for 1,261 cities; i.e., for 57.5 percent of all cities, the most important collaborator is the United States. The second-ranked United Kingdom is the most important collaborator for only nine percent of all cities. As a matter of fact, there are very few countries in the world (e.g., Northern African Arabic countries, Saudi Arabia, Portugal, Austria,

---

[10] According to Leydesdorff & Wagner (2009) the scientific output of the United States (and the European Union), in terms of the number of articles published annually, grows slowly and its world share has been decreasing for decades. Conversely, the scientific output of developing countries and China has grown very rapidly. When measuring only the top 1% most highly cited papers, the United States is still on top of the world, surpassing not only China, but also the European Union.

Switzerland, and Indonesia) in which the cities' most important collaborator is not the United States (Fig. 2).

**Table 4. Countries, as the most important collaborators**

| Rank | Country | Number of cities for which the most important collaborator is the given country* | Proportion of cities in the dataset | Number of co-authored articles |
|---|---|---|---|---|
| 1 | United States | 1261 | 57.52 | 2,921,057 |
| 2 | United Kingdom | 198 | 9.02 | 299,819 |
| 3 | Canada | 160 | 7.29 | 88,576 |
| 4 | Germany | 153 | 6.97 | 318,007 |
| 5 | China | 121 | 5.52 | 78,179 |
| 6 | France | 67 | 3.05 | 77,896 |
| 7 | Japan | 26 | 1.19 | 10,875 |
| 8 | Spain | 23 | 1.05 | 12,953 |
| 9 | Saudi Arabia | 21 | 0.96 | 18,777 |
| 10 | Italy | 18 | 0.82 | 4,185 |
| 11 | Australia | 17 | 0.77 | 3,371 |
| 12 | Russian Federation | 15 | 0.68 | 10,940 |
| 13 | South Korea | 15 | 0.68 | 2,581 |
| 14 | Egypt | 11 | 0.50 | 9,840 |
| 15 | Malaysia | 10 | 0.46 | 1,747 |

*In this context the most important collaborator for a given city is the country whose authors are represented in the greatest number of co-authored articles

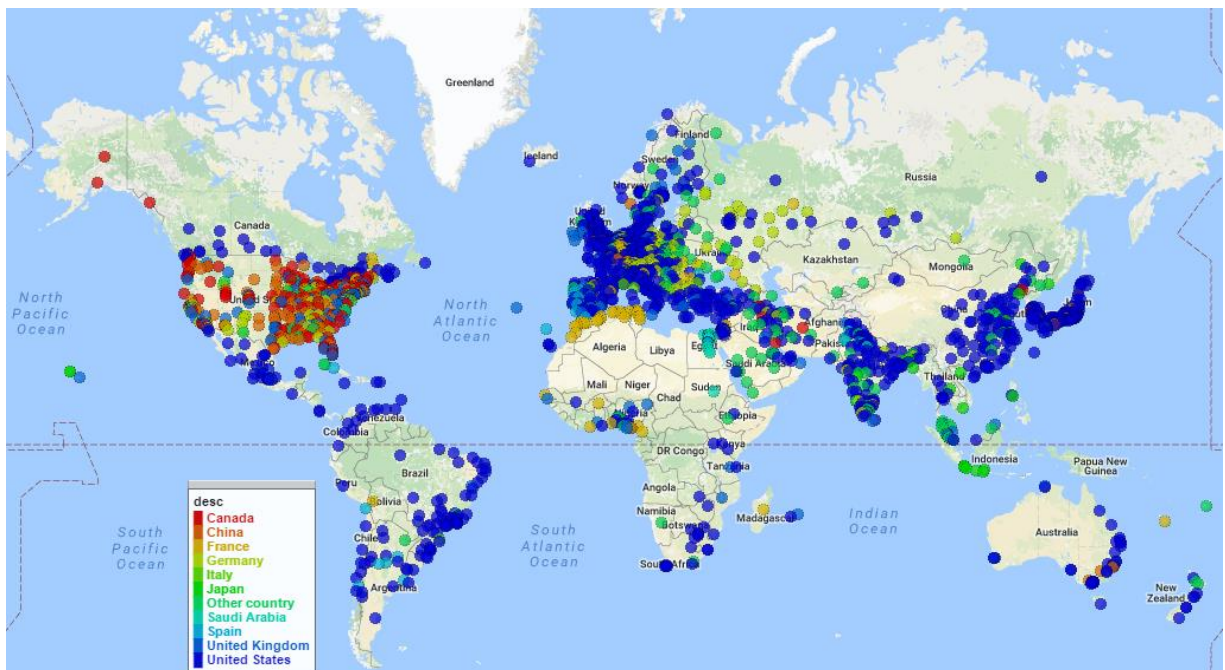

**Fig. 2. The most important collaborators for cities, 1986-2015**
**(A zoomable version of this map is available at https://dea.lib.unideb.hu/dea/handle/2437/241236)**

First, it is worth examining which countries are the most important collaborators for cities in the United States. Though a regular geographical pattern cannot be detected, there is evidence correlating the United Kingdom and the East Coast of the United States, especially in cities having the highest scientific output (e.g., New York City, Boston, Cambridge, Philadelphia, and Washington D.C.). The following associations with regions of the United States were similarly made: China is an important collaborator for cities in the midwestern and southern regions, and Germany is an important collaborator for cities in New Mexico and in the Chicago metropolitan area. In fact, 91.6 percent of all cities in the United States had approximately four countries associated as important collaborators. Canada is the most important collaborator for every third city (155) in the United States, followed by the United Kingdom (119 cities), China (102 cities), and Germany (62 cities). Behind every anomaly, specific reasons can be found. For example, the most important collaborator for Honolulu, Hawaii and Kaneohe, Hawaii is Japan, mostly because of joint research projects being conducted in the field of earth and planetary sciences.

In conclusion, in order to examine cities' international collaborations in a more realistic way, the distorting effect generated by the United States' hegemony in international science needs to disappear; i.e., the United States should be removed from the countries being examined. Table 5 shows that after removing the United States, the United Kingdom becomes the most important collaborator for the greatest number of cities, and the significance of Japan, Sweden, Spain, and Australia also increases.

**Table 5. Countries, as the most important collaborators, not including the United States**

| Rank | Country | Number of cities for which the most important collaborator is the given country, not including the United States* | Proportion of cities in the dataset, not including the United States | Number of co-authored articles |
|---|---|---|---|---|
| 1 | United Kingdom | 619 | 28.20 | 938,879 |
| 2 | Germany | 422 | 19.23 | 776,259 |
| 3 | China | 239 | 10.89 | 180,332 |
| 4 | Canada | 196 | 8.93 | 96,364 |
| 5 | France | 147 | 6.70 | 207,745 |
| 6 | Japan | 121 | 5.51 | 101,155 |
| 7 | Spain | 87 | 3.96 | 55,766 |
| 8 | Australia | 55 | 2.51 | 27,455 |
| 9 | South Korea | 39 | 1.78 | 4,813 |
| 10 | Italy | 33 | 1.50 | 11,993 |
| 11 | Saudi Arabia | 26 | 1.18 | 19,244 |
| 12 | Russian Federation | 21 | 0.96 | 12,924 |
| 13 | Sweden | 16 | 0.73 | 21,521 |
| 14 | Netherlands | 15 | 0.68 | 20,247 |
| 15 | Switzerland | 15 | 0.68 | 9,562 |

*In this context the most important collaborator for a given city is the country whose authors are represented in the greatest number of co-authored articles

Fig. 3 shows the characteristics of 2,194 cities' international collaborations after removing the United States from the dataset. According to Frame & Carpenter (1979); Hoekman, Frenken, & Tijssen (2010); Leclerc & Gagné (1994); Luukkonen, Persson, & Sivertsen (1992); and Nagpaul (2003), historical and linguistic features are the most significant determining factors in international scientific collaboration, a fact underpinned by this analysis.

Cities in Spanish-speaking nations in Latin America have Spain as their most important collaborator. However, Brazil, in which the official language is Portuguese, mainly collaborates with the United Kingdom (21 out of 60 Brazilian cities) and France (15 out of 60 cities).

Africa is more complex than Latin America; however, its historical and linguistic features (i.e., effects of European colonisation) determine international collaborations as well. For example, for many cities in Western African countries and for every city in the North African Arabic countries, France is the most important collaborator; however, for English-speaking countries of those regions, the United Kingdom is the most important collaborator. Egyptian cities follow a different pattern, as their most important collaborator is Saudi Arabia; and conversely, for Saudi Arabian cities, it is Egypt. For many cities in Israel, the most important collaborator is Germany. This is due to the fact that in 1986, the German-Israeli

Foundation for Scientific Research and Development was established to distribute funding for joint German-Israeli research (Zimmerman, Glänzel, & Bar-Ilan, 2009).

Iran, a country with one of the highest volumes of scientific output in Western Asia, shows the most complex picture, since its most important collaborators, the United Kingdom, Canada, and Australia are connected by the English language. Surprisingly, for the greatest number of Indian cities, the most important collaborator is not the former coloniser country (United Kingdom), but Germany (for 32 percent of all Indian cities). The reason for this is that there is a much greater overlap in joint research projects between India and Germany, primarily in the field of chemistry (which is India's most productive discipline) (Basu & Kumar, 2000). This is not the case in other disciplines, which decreases the likelihood of collaboration between India and the United Kingdom. Japan shows a pattern of being a regional rather than a global actor. This is reflected by the fact that Japan is the most important collaborator to a total of 121 cities, but only six are outside Asia. Cities with the highest scientific output in leading East Asian economies (i.e., China, Japan, and South Korea) generally collaborate with countries within the same region. For example, for 44 out of 110 Chinese cities, it is Japan and for 20 out of the Chinese cities, it is Australia; for 94 out of 151 Japanese cities, it is China; and for 30 out of 39 South Korean, it is also China who is the most important collaborator. Furthermore, Japan is the most important collaborator for the greatest number of cities in Bangladesh, Thailand, and Indonesia. For Australian cities that produce the highest scientific output (i.e., Sydney, Melbourne, Brisbane, Perth, Adelaide, and Canberra), the United Kingdom is the most important collaborator; however, for smaller cities in Australia, it is China. Cities in New Zealand most intensively collaborate with Australia.

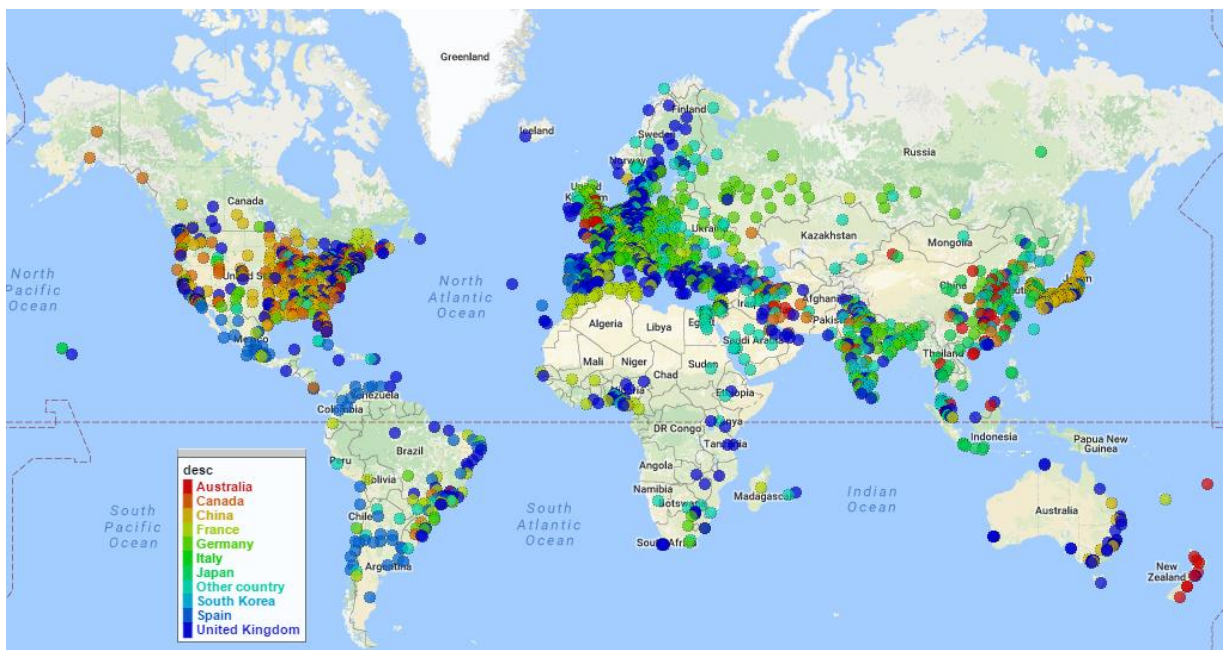

**Fig. 3. The most important collaborators for cities, not including the United States, 1986-2015**
**(A zoomable version of this map is available at https://dea.lib.unideb.hu/dea/handle/2437/241236)**

Europe shows the most complex picture, because in European countries, linguistic ties significantly influence international collaborations in science, with dominance of English (Almeida, Pais, & Formosinho, 2009; Thelwall, Tang, & Price, 2003). The two main scientific actors in Europe are Germany and the United Kingdom. The former is the most important collaborator for 422 cities worldwide, including 256 European cities; and the latter is for 619 cities worldwide, including 255 cities in Europe. This means that Germany and the United Kingdom are the most important collaborators for 67.9 percent of all European cities. As Fig. 3 shows, the most important collaborator for many cities in the United Kingdom is Germany, and vice versa. In addition to these two countries, only France is significantly involved with international collaboration in Europe. France is the most important collaborator for only 70 European cities; i.e., for 9.3 percent of all cities in Europe. For French cities, the most important collaborators are Germany and the United Kingdom. Naturally, the most important collaborator for Portuguese cities is Spain, but for Spanish cities (including Madrid and Barcelona, which produce the nation's highest scientific output), it is the United Kingdom. Cities located in the northeast part of Spain (e.g., Girona and Blanes) most intensively collaborate with France. The same phenomenon can be seen in France: the most important collaborator for most cities in southern France is Italy; but for cities in the Great East region, it is Germany. In these latter cases, geographical proximity is a more important determinant in scientific collaboration than are linguistic ties. In multi-language countries, significant correlations can be seen between the geographical pattern of international collaborations and the main languages. In Belgium, the most important collaborator for cities in the Dutch-speaking Flemish Region (e.g., Leuven and Ghent) is the Netherlands, but for cities in the French-speaking Walloon Region (e.g., Liège and Louvain-la-Neuve), it is France. Switzerland is more complex as compared to Belgium, because the most important collaborator for cities in the western part of Switzerland (e.g., Geneva, Lausanne, and Neuchâtel) is France; for cities in the northern cantons (e.g., Zurich, Basel, and Bern), it is Germany; and in cities located near the Italian border (Bellinzona and Lugano), the most intensive collaboration is with Italy. For the majority of central and Eastern European cities, the most important collaborator is Germany; however, some exceptions do occur. For example, in every Slovakian city, the most important collaborator is the Czech Republic; some of the Silesian cities most intensively collaborate with Ukraine; and the most important collaborator for half of the Romanian cities is France. In the case of Czech and Slovakian cities, the following reasons can be linked to the collaboration patterns. Between 1918 and 1993, the Czech Republic and Slovakia constituted Czechoslovakia, a country that was part of the Eastern European communist bloc. As parts of Czechoslovakia, the Czech Republic and Slovakia shared a common higher education system, and research activities were conducted under the direct control of the government (Koucký, 1990). In 1993, after the breakdown of the communist regime, Czechoslovakia split into the two sovereign states of the Czech Republic and Slovakia. Given the long common history of these countries, scientific collaboration ties between Czech and Slovakian researchers have remained strong; however, the international collaboration patterns of the Czech Republic have radically changed, in that Western European countries (e.g., Germany, France, and the United Kingdom) have become the country's most important collaborators (Almeida, Pais, & Formosinho, 2009). By contrast, the most important collaborators for Slovakia have remained its neighbouring countries (e.g., the Czech Republic, Poland, and Hungary) (Kozak, Bornmann, & Leydesdorff, 2014).

In Northern European countries (i.e., Denmark, Finland, Iceland, Norway, and Sweden), 36 out of 69 cities most intensively collaborate with the United Kingdom; but, amongst one another, the most important collaborator is Sweden (for 14 out of 69 cities).

Finally, the Russian Federation's most important collaborator is Germany (Almeida, Pais, & Formosinho, 2009; Wilson & Markusova, 2004), so it is not surprising that for the majority of Russian cities (more precisely, for 28 out of 39 cities), it is Germany, as well.

Several cities appear amongst the exceptions that most intensively collaborate with neighbouring or geographically close countries, especially in the field of earth and planetary sciences (e.g., Apatity and Petrozavodsk: Finland; Yakutsk and Vladivostok: Japan). Among the former member-states of the Soviet Union (i.e., Armenia, Belarus, Georgia, Kazakhstan, Ukraine, and Uzbekistan), cities share quite intensive scientific connections to Germany; however, their most important collaborator is the Russian Federation, of course. This pattern is not representative of Baltic countries. This is partly because the most important collaborator for cities in Latvia and Lithuania is Germany[11], but Estonian cities (Tartu and Tallinn) most intensively collaborate with Finland, since both Estonian and Finnish languages are members of the Finnish branch of the Uralic language family (Petersoo, 2007).

In conclusion, without the distorting effect of the United States, historical (e.g., colonisation) and linguistic features determine cities' international collaborations. Results show that the United Kingdom, Germany, and China, and to a lesser extent, France, Japan, and Spain, are the main international scientific collaborators.

**3.3. The most productive disciplines**

Scopus classifies scientific publications into 27 disciplines (subject areas). According to the content of the Scopus database, the greatest number of publications, including journal articles, belong to the field of medicine[12], which means that medicine is the most productive discipline in the world. Therefore, it is not surprising that medicine is the most productive discipline in 44.7 percent in all cities (Table 6 and Fig. 4). There is, however, an important reason why medicine is so geographically dispersed worldwide: Medical research is conducted not only in medical universities, which are generally concentrated in large cities and university towns, but also in hospitals, which can be found in most cities in the world.

Furthermore, the following correlation can be observed between cities' total scientific output and the significance of medicine: The higher a city's scientific output, the more likely it is that the greatest number of articles are published in the field of medicine. In 55 out of 73 cities having more than 100,000 articles (75.3 percent of this group), medicine is the most productive discipline.

**Table 6. The most productive disciplines in cities, 1986-2015**

| Rank | Disciplines | Number of cities in which the given discipline is the most productive | Number of articles |
|---|---|---|---|
| 1 | Medicine | 981 | 9,531,147 |
| 2 | Physics and Astronomy | 309 | 1,814,992 |
| 3 | Agricultural and Biological Sciences | 290 | 655,898 |
| 4 | Engineering | 180 | 1,166,999 |
| 5 | Chemistry | 106 | 206,997 |
| 6 | Social Sciences | 85 | 82,096 |

---

[11] German colonists in the 12th and 13th centuries settled in numerous cities on and near the east Baltic coast, such as Reval (Tallinn), Riga, and Dorpat (Tartu), which became members of the Hanseatic League. This commercial confederation grew out of North German towns and was dominated by Lübeck (Dollinger, 2000). In the 16th century, the League lost its importance in the Baltic region; however, the German historical and cultural legacy significantly influenced the modern Baltic States.

[12] It is necessary to note that Scopus includes MEDLINE, a freely available dataset compiled by the United States National Library of Medicine and searchable via the platform PubMed. MEDLINE indexes more than 5,600 biomedical journals. About 60 percent of all journals indexed by Scopus are classified in the Health Sciences subject cluster (Elsevier, 2016). However, it is not only Scopus, but also Web of Science that is biased toward biomedicine.

| 7 | Biochemistry, Genetics and Molecular Biology | 62 | 169,555 |
|---|---|---|---|
| 8 | Materials Science | 61 | 105,558 |
| 9 | Earth and Planetary Sciences | 57 | 170,592 |
| 10 | Environmental Science | 16 | 25,366 |
| 11 | Veterinary | 13 | 14,620 |
| 12 | Mathematics | 10 | 6,023 |
| 13 | Pharmacology, Toxicology and Pharmaceutics | 8 | 5,469 |
| 14 | Computer Science | 7 | 9,262 |
| 15 | Chemical Engineering | 3 | 3,016 |
| 16 | Immunology and Microbiology | 3 | 3,998 |
| 17 | Energy | 2 | 2,105 |
| 18 | Business, Management and Accounting | 1 | 788 |
| 19 | Economics, Econometrics and Finance | 1 | 690 |
| 20 | Psychology | 1 | 534 |

Contrary to the distribution of the most important international collaborators, the distribution of the most productive disciplines follows a well-defined geographical pattern. Table 6 shows that although medicine is the most productive discipline in the majority of cities, regional divergences can be observed in both countries and continents. For example, medicine is the most productive discipline in 376 European cities; i.e., in half of all cities on the continent, but the proportion of cities in which medicine is the most productive discipline is the highest in Oceania (59.3 percent) and Africa (55.5 percent). There are only 20 cities in China in which medicine is the most productive discipline and 41 cities for which engineering is the most productive. In India, chemistry is the most productive discipline out of the majority of cities (32 cities), but medicine is the most productive in only 19 cities. Furthermore, medicine is the most productive discipline in 56.3 percent of Japanese cities, and in 42.9 percent of US cities.

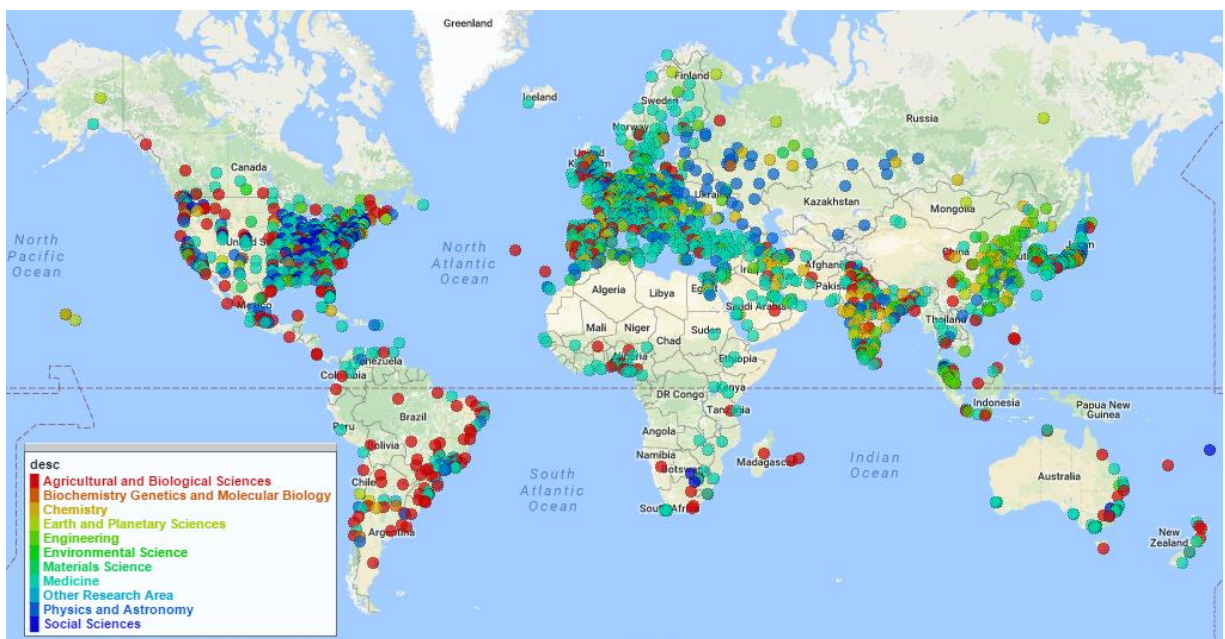

**Fig. 4. The most productive disciplines in cities, 1986-2015**
**(A zoomable version of this map is available at https://dea.lib.unideb.hu/dea/handle/2437/241236)**

In 69 out of 139 cities in Latin America, the greatest number of articles were created in the field of agricultural and biological sciences, since the first significant scientific developments were introduced in this discipline in Latin America (Vessuri, 1995). Notable exceptions are major cities located in Rio de Janeiro and São Paulo; moreover, cities in Colombia and Venezuela can count medicine as the most productive discipline. Although agricultural and biological sciences are the most productive disciplines in most Mexican cities, physics, astronomy, and medicine are also important research fields in several cities, including Mexico City, the city where the greatest number of articles are created in Mexico.

Considerable regional differences exist between major East Asian countries (i.e., China, Japan, and India) that produce the highest scientific output. In 37.3 percent of Chinese cities, the most productive discipline is engineering (see Fig. 4. and Fig. 5), while medicine and chemistry are the most productive disciplines in 20 Chinese cities, respectively. In Beijing, the city with the highest scientific output in the world, 228,537 articles were created in the field of engineering in the period between 1986 and 2015, more than were produced in Los Angeles during this period. Medicine is a marginal discipline in Beijing, as it is surpassed by physics and astronomy, materials science, and chemistry in terms of the number of articles. In India, chemistry is the most productive discipline in 32 cities. This is followed by engineering, in terms of the number of cities (21 cities). Agricultural and biological sciences seem to be the most productive disciplines in cities in the northern states (Chandigarh, Haryana, and Uttarakhand). In South Indian states (e.g., Tamil Nadu), engineering is the most productive discipline, and in the mid-regions of the country, the greatest number of articles are published in chemistry. In Mumbai, Kolkata, and Bangalore; i.e., in cities having the highest scientific output, physics and astronomy are the most productive disciplines. Meanwhile in New Delhi, it is medicine. Finally, in Chennai the greatest number of articles are created in engineering. The most productive discipline in many Japanese cities is medicine; moreover, biochemistry, genetics and molecular biology is also a very productive discipline. In half of South Korean cities, including Seoul, medicine is the most productive discipline, while in one-third of all cities, engineering and material sciences are the most productive disciplines.

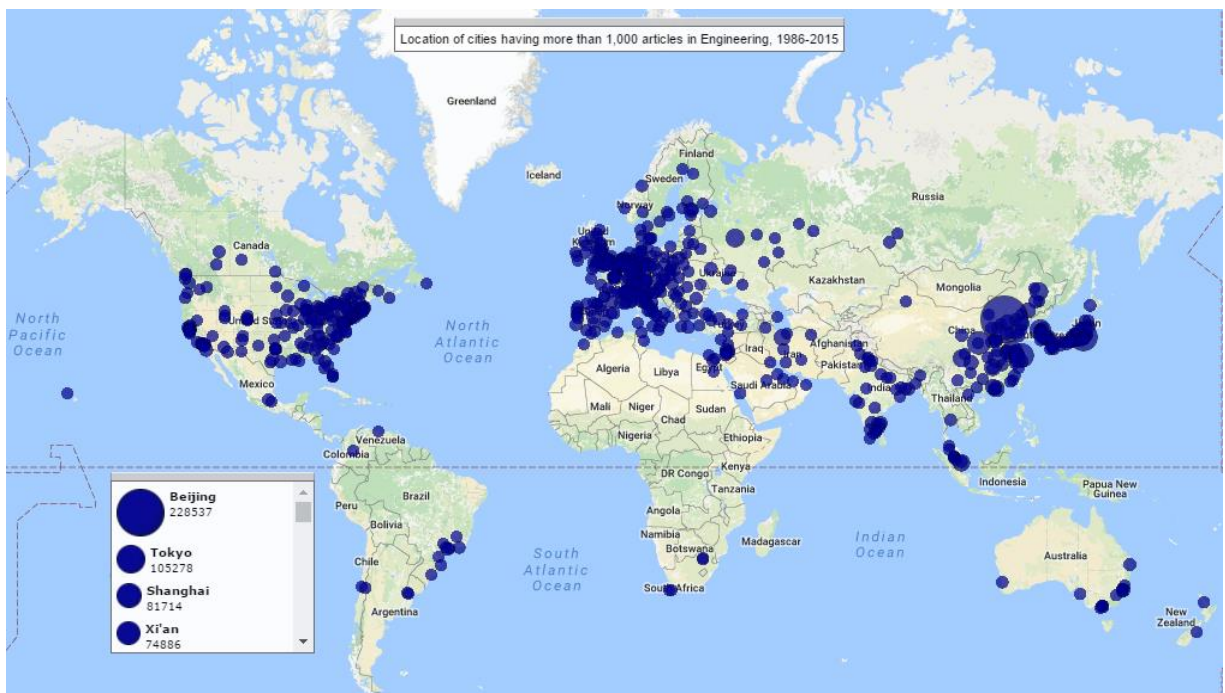

**Fig. 5. Location of cities having more than 1,000 articles in Engineering, 1986-2015**
**(A zoomable version of this map is available at https://dea.lib.unideb.hu/dea/handle/2437/241236)**

In 376 out of 752 European cities, medicine is the most productive discipline, followed by physics and astronomy (132 cities), and agricultural and biological sciences (69 cities). The geographical distribution of cities with the greatest number of articles is unequal in the fields of medicine and physics and astronomy. For example, in central and Eastern European cities, physics and astronomy has a greater importance, since it is the most productive discipline in 47 out of 118 cities, while medicine is the most productive discipline in only 21 cities. Naturally, the greatest number of cities in which agricultural and biological sciences is the most productive discipline can be found in countries that are leading agricultural producers (e.g., France, Spain, Portugal, and some Central and Easter European countries). However, agricultural and biological sciences is also an important discipline in the United Kingdom; in fact, the most productive discipline in 10 cities.

It is not surprising, that in the greatest number of Northern American cities, medicine is the most productive discipline, but to a lesser extent than in Europe. Medicine is the most productive in many cities that have more than 100,000 articles, except for Cambridge, Massachusetts and Berkeley, California, where physics and astronomy is the most productive field, because these cities are home to prestigious universities (e.g., MIT and UC Berkeley) that are traditionally involved in research in physical sciences and engineering. In Canadian cities with the highest scientific output (i.e., Toronto, Montreal, Ottawa, Vancouver, Edmonton, Hamilton, and Calgary), medicine is the most productive discipline, but in small and mid-sized cites (i.e., in 25.5 percent of all cities) it is the agricultural and biological sciences. In the United States, social sciences is the most productive discipline in the second greatest number of cities, and this fact makes the United States unique: in fact, 82.4 percent of all cities in the world that count social sciences as the most productive discipline are in the United States (see Fig. 4). However, the city that has the highest scientific output in social sciences is London, having produced 46,016 articles in the field during the study period. In this discipline, London is followed by four US cities: New York City, Washington DC, Chicago, and Los Angeles. Fig. 6 shows that the scientific output of East Asian cities is generally quite low. For example, only 1.7 percent of all articles created in Beijing belong to the field of social sciences (though interestingly, Beijing has the eighth-highest scientific output in social sciences in the world). According to Zhou, Thijs, & Glänzel (2009b), there are many reasons why China (and therefore, Chinese cities) have "low international visibility" in social sciences. The reasons include the following: 1) In China, the attribution of national orientation on social sciences is less favourable than that of natural sciences; 2) there is a perceptible influence of the official political ideology on social sciences; 3) the separated administration systems for natural sciences and social sciences may hinder collaboration between the two fields; and finally, 4) due to the special evaluation system, there are no measures to stimulate social scientists to publish internationally; i.e., in journals included in the SCIE/SSCI (Zhou, Thijs, & Glänzel, 2009b). However, as can be seen in Fig. 6, the low international visibility in social sciences also applies to Seoul and Tokyo, which suggests that some of the aforementioned factors characterise not only China, but East Asia in general. For example, a smaller number of articles were created in social sciences in Tokyo compared to Glasgow; however, the total amount of scientific output of Tokyo is 7.7 times greater than that of Glasgow.

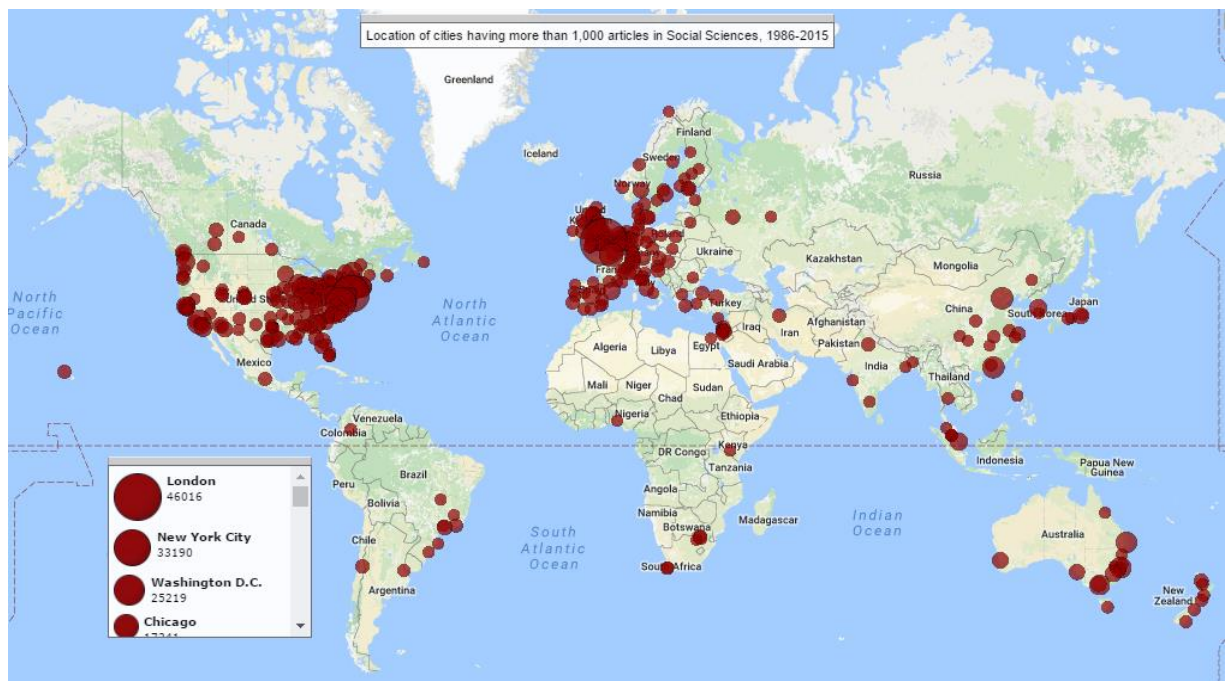

**Fig. 6. Location of cities having more than 1,000 articles in Social Sciences, 1986-2015**
(A zoomable version of this map is available at https://dea.lib.unideb.hu/dea/handle/2437/241236)

In Australia and New Zealand, the most productive disciplines in terms of the number of cities are medicine and agricultural and biological sciences; however, in cities having the highest scientific output, medicine is the most productive. The most notable exception is Canberra, where the greatest number of articles were created in the field of physics and astronomy.

## 4. Discussion

This analysis gives information on the total scientific output of cities worldwide, and the change in the output over time (by decades). Furthermore, it has been presented that for a given city (more precisely, authors affiliated with that city), which countries (more precisely, authors affiliated with those countries) are the most important collaborators, and which disciplines are the most productive in each city. However, the results make it clear that the analysis has certain limitations, and for this reason, further targeted research is needed to improve the results. Some limitations and their possible solutions are summarised in the following:

**4.1. Total publication output vs. output of highly cited papers**

In this paper, the scientific output of a given city was obtained by measuring the total number of journal articles produced in that city in the period from 1986 to 2015. Results show that the scientific output of Chinese cities has significantly increased in 30 years, especially in the last decade (the period from 2006 to 2015). It is, however, important to investigate differences among cities worldwide in terms of their scientific output and citations. In the past few years, multiple studies have been written focusing on cities' or urban regions' output of highly cited papers (e.g., Bornmann & Leydesdorff, 2011; Bornmann, Leydesdorff, Walch-Solimena, & Ettl, 2011; Bornmann & Waltman, 2011; and Bornmann & Leydesdorff, 2012). These studies clearly show (especially in comparison with the results of this analysis) that cities having high publication output do not necessarily have a high output of highly cited papers. For example, Zhou, Thijs, & Glänzel (2009a) found that despite having high scientific output in terms of the number of publications, the citation impact of cities in mainland China (except for that of Hong Kong) was much lower than the world average. This result has also been reinforced by the work of Andersson, Gunessee, Matthiessen, & Find (2014: 2969), who stated that "even Beijing lacks the impact of Western cities with similar publication volumes, such as London or Paris".

Therefore, future research should deal with not only cities' total scientific output but also the output of highly cited papers; i.e., not only quantity, but also quality. Another important issue should be the comparison of the scientific output with some other relevant data; e.g., the population or the size of the city.

**4.2. Challenges in comparing cities at the international level**

In this analysis, articles have been assigned to cities as reported by authors of the articles and indicated as such in the Scopus database (that is why the Moffett Federal Airfield and the Research Triangle Park have been listed as cities). However, cities worldwide are quite different territorial units in terms of population and area, which makes them difficult to compare. In Northern America, Western Europe, Australia, and Japan (e.g., New York, Los Angeles, Boston, London, Paris, Madrid, Sydney, Melbourne, Tokyo, and Osaka), major cities have large metropolitan areas in which many autonomous villages, towns, and small and medium-sized cities are located, and some of them have high scientific output (see, for example, Grossetti, M., Eckert, D., Gingras, Y., Jégou, L., Larivière, V., & Milard, B., 2014). Conversely, Chinese cities, especially the prefecture-level cities (also direct-controlled municipalities), do not have "metropolitan areas" in the usual sense of the term. The prefectural-level city (in this analysis, every Chinese city is a prefectural-level city) is often not simply a city in the usual sense of the term, but instead an administrative unit comprising a main central urban area (i.e., the built-up area of the prefectural-level city), and its much larger surrounding rural area containing counties, many smaller cities, towns and villages. Chinese cities (municipalities) in terms of their size cannot be comparable with any other cities (municipalities) in the world (Tödtling, 1994; van Noorden, 2010). For example, Boston is generally considered to be one of the world's best cities for science, and is among the leading cities on the basis of its scientific output (in this analysis, Boston is ranked 9[th] in terms of the total scientific output). Yet, regarding size, Boston (i.e., the city of Boston), with a population of 673,000 and an area of 232 km$^2$, is incomparable with Beijing, a Chinese (direct-controlled) municipality with a population of 21.7 million and (by comprising 16 urban, suburban, and rural districts) an area of 16,000 km$^2$. However, Beijing and all other Chinese cities in terms of their population and area, are comparable with metropolitan areas as they are defined in the United States, Europe, Japan, Australia, and so on. For example, the Greater Boston Area (generally referred to as simply Boston) is a metropolitan area[13] with a population of 8.2 million, and its area covers more than 25,000 km$^2$. The Greater Boston Area produces one of the world's highest scientific output (it produced more than 700,000 articles in the period from 1986 to 2015), but only 45.7 percent of the total number of articles comes from Boston (i.e., from the central city), More than 50 percent of the publications come from suburban settlements (including 27.3 percent from Cambridge, which borders Boston).

---

[13] The Greater Boston Area corresponds to the Boston-Worcester-Providence, MA-RI-NH-CT Combined Statistical Area as defined by the Office of Management and Budget.

In conclusion, it is important to merge cities having markedly different populations and areas into metropolitan areas that can be comparable territorial units in terms of size on the international level. Naturally, there are multiple alternatives for combining local territorial units into metropolitan areas. On the one hand, most national and international statistical agencies have already defined metropolitan areas, such as the ESPON in the European Union (functional urban areas), the Office of Management and Budget in the United States (metropolitan statistical areas and combined statistical areas), and the Statistics Bureau of Japan (major metropolitan areas and metropolitan areas). These larger territorial units can be more effectively used for spatial scientometrics analyses than simply using cities. On the other hand, some researchers have introduced territorial demarcation methods, which can be used specifically for spatial scientometrics (Matthiessen & Schwarz, 1999; Maisonobe, Eckert, Grossetti, Jégou, & Milard, 2016). The analysis performed in this study should be repeated again measuring on the level of metropolitan areas.

**4.3. Challenges in mapping cities' international collaboration**

In this paper, for a given city, only the most important collaborator country has been presented; i.e., only the country in which the largest number of co-authors are located. Results clearly show the hegemony of the United States in international science, and the significance of the United Kingdom, Germany, and China as important collaborator countries. However, when outlining the future direction of research on international collaboration of cities, three factors should be taken into account.

First, the difference between the main collaborator of a given city and the second-ranked (even the third-ranked) collaborator is frequently very small in terms of the number of articles. For example, the most important collaborator of Boston is the United Kingdom, with 16,638 co-authored articles, but Germany, the second-ranked collaborator (with 16,562 co-authored articles), closely follows the United Kingdom, representing a difference between the two countries of less than 0.5 percent (76 articles). This case is not unique; in fact, there are many cities worldwide in which the number of co-authored articles of the most important collaborator exceeds that of the second-ranked collaborator by less than a 1 or 2 percent margin. In such cases, the position of the "most important" collaborator is uncertain, and because of this uncertainty, the measurement of international collaboration of cities should be revised.

Second, it should be examined what reasons lie behind the instances in which the most important collaborators of certain cities in a given country are markedly different from the most important collaborator of that country. In this paper, it has been pointed out that the geographical proximity and the linguistic, cultural, and historical ties fundamentally influence the international collaboration patterns of cities; however, there are some cases that are difficult to explain. For example, not surprisingly, the most important collaborator of France is the United States, as is true for most French cities. There are some rational differences from this pattern (e.g., the most important collaborator of Mulhouse, a city located only 10 km from the German border, is Germany); however, in some cases, the international collaboration patterns cannot be explained by geographical proximity nor by any relationships. The most important collaborator of Arras and Valenciennes, two northern French cities near the Belgian border, is Algeria, but the most important collaborator for every southern French city is a different country, despite the fact that they have closer geographical proximity, and some of them (e.g., Montpellier and Marseille) have stronger historical ties with Algeria. Troyes and Belfort are other remarkable examples, as their most important collaborator is China. It will be important to examine these cases one by one, in order to obtain explanations for the differences.

Third, the collaboration intensity of many cities worldwide, especially that of Chinese cities, is quite weak. For example, in the period from 1986 to 2015, more than 100,000 articles were produced in Changsha, but the total number of collaborations is less than 17,000; i.e., the value of collaboration intensity of Changsha is 0.17 per article. Contrary to Changsha, the value of collaboration intensity of Amsterdam is 1.2 per article, because, in the period from 1986 to 2015, 150,000 articles were produced in Amsterdam, and the number of its collaborations exceeds 180,000. That is, despite having high scientific output, some cities' collaboration intensity is very weak, and for this reason it should be considered to introduce a reasonable threshold when mapping cities' international collaboration.

**4.4. Challenges in mapping the productivity of disciplines in cities**

Regarding the productivity of disciplines, in some cities, the main disciplines are close to each other in terms of the number of articles. However, the difference between the most productive disciplines (i.e., between the first and the second-/third-ranked disciplines) is generally much larger than that between the most important collaborators (i.e., between the first and the second-/third-ranked collaborators). The reason for this is that authors affiliated with a given city can collaborate with co-authors from more than 200 countries in the world, but Scopus classifies scientific publications into 27 disciplines only. For example, the most productive discipline in Boston is medicine, comprising 38.6 percent of all articles, the second-ranked discipline is biochemistry, genetics and molecular biology with 17.8 percent, and neuroscience is third with 5.1 percent. The difference between the most productive discipline and the second-ranked discipline is more than 20 percent, while the difference between Boston's most important collaborator (the United Kingdom) and the second ranked collaborator (Germany) is less than 1 percent. It would be far more interesting to examine whether there is a difference between the most productive discipline of a country and that of cities located in that country. For example, out of 43 Australian cities included in this analysis, medicine is the most productive discipline in 28 cities; i.e., in 65 percent of all cities, which is above the world average. However, there are some cities in which medicine is the most productive discipline, but those cities most intensively collaborate with a country in which the most productive discipline is not medicine. In Wollongong, for example, medicine is the most productive discipline, but its most important collaborator is China, a country in which the most productive discipline is engineering. A more thorough analysis shows that the collaboration between authors affiliated with Wollongong and co-authors located in China is strong in materials science, engineering, and physics and astronomy; i.e., disciplines that are very productive in China. When examining articles in the field of medicine, it is discovered that authors affiliated with Wollongong most intensively collaborate with co-authors located in the United States and the United Kingdom, while China is only 5[th] in this ranking. The case of Wollongong is not unique, implying that such specific cases should be examined again one by one.

**4.5. Changing collaboration patterns through time**

It should be reviewed how the collaboration patterns of cities have changed over time. In most cases, in the period from 1986 to 2015, what did not change was which country was the most important (or even the second or the third most important) collaborator of a city; however, there are some cities in which significant changes could be detected. For example, in West Lafayette (home to Purdue University), engineering was the most productive discipline in each decade. As for international collaboration, between 1986 and 1995, the most important collaborator of West Lafayette was Japan (surpassing the second-ranked Italy by only a few articles), then in the next decade it was Canada (surpassing the second-ranked Germany by only a few articles), and between 2006 and 2015, China became West Lafayette's most important collaborator (significantly surpassing the second-ranked South Korea). The increasing importance of China in international science has been reinforced by many other cases. It is also an interesting example that between 1986 and 1995, and between 2006 and 2015, the United Kingdom was the most important collaborator of both Boston and New York, but between 1996 and 2005 (i.e., the middle decade), Boston most intensively collaborated with Germany, and New York with Japan. So, the question is, why did the collaboration intensity between US cities and the United Kingdom decrease in the period from 1996 to 2005, or were the cases of Boston and New York unique? All of these examples suggest that deeper analysis is required.

## 5. Conclusion

In this paper, I performed a spatial scientometric analysis to investigate cities' scientific output. I calculated a given city's scientific output based on the total amount of journal articles produced by authors affiliated with a city. In the period between 1986 and 2015, 2,194 cities had at least 1,000 articles indexed by Scopus. The greatest number of cities were located in the United States, Europe, and Japan, with an increasing number of cities in China and India.

Results show that major world cities (e.g., Tokyo, London, New York City, Paris, and Moscow) produce the highest scientific output; however, the gap between these world cities and some Chinese cities has been gradually closing. Furthermore, some emerging cities (e.g., Seoul, São Paulo, and Tehran) have been identified as having growing scientific output. Beijing had the highest scientific output in the world from 2006 to 2015, since more articles were created in Beijing than second-ranked Tokyo produced in 30 years. In the period from 2006 to 2015, the number of scientific articles published in Chinese cities has grown rapidly, and it is predicted that this trend will continue into the future. Furthermore, Chinese cities' world share of scientific output will also likely increase.

The hegemony of the United States in international science is well-known, and is reinforced by the fact that for 1,621 cities; i.e., for 73.5 percent of all cities, the most important collaborator in terms of the number of co-authored articles is the United States. The United Kingdom is the second-ranked country and is the most important collaborator for only nine percent of all cities in the word. After removing the United States from the list of examined countries (i.e., after eliminating the distorting effect generated by the United States' hegemony in international science), it is discovered that cities' most important collaborators are the United Kingdom, Germany, and, since the beginning of the 2000s, China. Furthermore, it is identified that cities' international collaborations are primarily determined by historical, linguistic, and regional features.

Results show that medicine is the most productive discipline in the greatest number of cities in the world; however, there are unique geographical patterns that can be observed. For example, agricultural and biological sciences is the most productive discipline in a majority of Latin American cities, while in many Indian cities, chemistry is the most productive. Many cities in the United States and in the United Kingdom are characterised by high output in social sciences, and in China, the greatest number of articles were created in engineering.

Finally, it should be noted that the analysis has certain limitations regarding cities' territorial demarcation, cities' international collaboration, and the productivity of disciplines. However, this analysis provides a good basis for future work that will bring us closer to understanding how cities are engaged in international science.

**Acknowledgement**

This paper is supported by the János Bolyai Research Scholarship of the Hungarian Academy of Sciences.

**Research data is available at**
https://dataverse.harvard.edu/dataset.xhtml?persistentId=doi:10.7910/DVN/VPRTEA